\PassOptionsToPackage{unicode}{hyperref}
\PassOptionsToPackage{hyphens}{url}
\documentclass[
]{article}

\usepackage{amsmath,amssymb,mathabx,commath}
\usepackage{float}

\usepackage{tikz}
\usetikzlibrary{fit, positioning, shapes.geometric, decorations.pathreplacing, calc, arrows.meta, positioning}
\usepackage{tikz-cd}

\tikzset{stretch/.initial=1}
\newcommand\drawloop[4][]%
{\draw[shorten <=0pt, shorten >=0pt,#1]
($(#2)!\pgfkeysvalueof{/tikz/stretch}!(#2.#3)$)
let \p1=($(#2.center)!\pgfkeysvalueof{/tikz/stretch}!(#2.north)-(#2)$),
    \n1= {veclen(\x1,\y1)*sin(0.5*(#4-#3))/sin(0.5*(180-#4+#3))}
    in arc [start angle={#3-90}, end angle={#4+90}, radius=\n1]%
}

\usepackage[margin=1.5in]{geometry}

\makeatletter
\@ifundefined{KOMAClassName}{
    \IfFileExists{parskip.sty}{%
        \usepackage{parskip}
    }{
        \setlength{\parindent}{0pt}
        \setlength{\parskip}{6pt plus 2pt minus 1pt}}
}{
    \KOMAoptions{parskip=half}}
\makeatother

\usepackage{longtable,booktabs,array}
\usepackage{calc} 
\usepackage{subcaption}
\usepackage{graphicx}
\makeatletter
\def\fps@figure{htbp}
\makeatother
\usepackage{adjustbox}

\makeatother

\usepackage{hyperref}
\hypersetup{
    pdftitle={Qualia \& Natural Selection},
    pdfauthor={Ryan Williams},
    hidelinks}

\usepackage{url}
\usepackage{natbib}
\usepackage{mathtools}
\usepackage{amsfonts}

\DeclareMathOperator*{\expectation}{\mathbb{E}}
\DeclareMathOperator*{\Var}{\mathrm{Var}}
\DeclareMathOperator*{\Cov}{\mathrm{Cov}}
\newcommand{\appropto}{\mathrel{\vcenter{
    \offinterlineskip\halign{\hfil$##$\cr
    \propto\cr\noalign{\kern2pt}\sim\cr\noalign{\kern-2pt}}}}}

\title{Qualia \& Natural Selection}
\makeatletter
\providecommand{\subtitle}[1]{
    \apptocmd{\@title}{\par {\large #1 \par}}{}{}
}
\makeatother
\subtitle{Formal Constraints on the Evolution of Consciousness}
\author{ Ryan
Williams\vspace{0.05in} \\ \newline\normalsize\url{ryan@cognitivemechanics.org} }
\date{April 2025}

\begin{document}

    \newgeometry{margin=1in}
    \thispagestyle{empty}
    \pagenumbering{gobble}

    \begin{center}
    {\Large\bfseries One-Page Summary: Qualia \& Natural Selection\\[4pt]
    \normalsize Formal Constraints on the Evolution of Consciousness}\\[6pt]

    \vspace{7pt}

    Ryan Williams\\
    \texttt{ryan@cognitivemechanics.org}
    \end{center}

    \vspace{7pt}

    \thispagestyle{empty}

    \textbf{Overview.} This paper establishes new quantitative methods for investigating the evolution of qualia.
    Using information-theoretic measures and the
    Price Equation, it derives a novel criterion—called the
    \emph{Qualitative Selection Limit} (QSL)—for evaluating whether natural selection on a physical system may be
    effectively transmitted into a qualitative domain.

    \vspace{7pt}

    \textbf{Setup.} We define a mapping \( \phi: S \rightarrow Q \), from structure $S$ to qualitative space $Q$, with a fidelity measure
    \[
        \hat{s} = \eta_{q \mid s} = \sqrt{\frac{\mathrm{Var}(\mathbb{E}[\phi(s)])}{\mathrm{Var}(q)}.}
    \]
    This fidelity expresses how determinable qualitative traits are from knowledge of structural traits.

    \vspace{7pt}

    \textbf{Price Equation for Qualia.} The Price Equation, applied to qualia can be written as:
    \[
        \Delta \bar{q} = \mathrm{Cov}(w_i, q_i) + \mathbb{E}[w_i \Delta q_i].
    \]
    By decomposing \( q = \bar{\phi}(s) + \epsilon \), with $\bar{\phi}(s)$ representing the portion of $q$ that can
    be determined via correlation with $s$, we show that:
    \[
        |\rho_{w, q}| \leq \hat{s} \cdot |\rho_{w, s}|,
    \]
    where $\rho$ is a the Pearson correlation coefficient.
    In other words, selective pressures on \( q \) are bounded by its fidelity to the domain of selection.

    \vspace{7pt}

    \textbf{Qualitative Selection Limit (QSL).}
    Taking into account other transmission effects from the Price Equation, selection becomes ineffective in the qualitative domain when:
    \[
        \frac{\mathrm{mutation}}{\mathrm{selection}} \gg \mathrm{fidelity} \quad \text{ i.e. } \quad \left| \frac{\mathbb{E}[w_i \Delta q_i]}{\sqrt{\mathrm{Var}(w_i)\mathrm{Var}(q_i)}} \right| \gg \hat{s}.
    \]
    This yields a natural threshold: low-fidelity mappings prevent selection from shaping qualitative traits.

    \vspace{7pt}

    \textbf{Simulation Results.} Experiments confirm QSL behavior across generations: low-fidelity systems show divergence between \( s \) and \( q \), even under selection.
    High noise in \( \phi \) (i.e. $\hat{s} \approx 0$) causes drift in quality $q$ despite convergence between the structure $s$ and fitness $w$.

    \begin{figure}[H]
        \centering

        \begin{subfigure}[t]{0.23\textwidth}
            \centering
            \includegraphics[width=\textwidth]{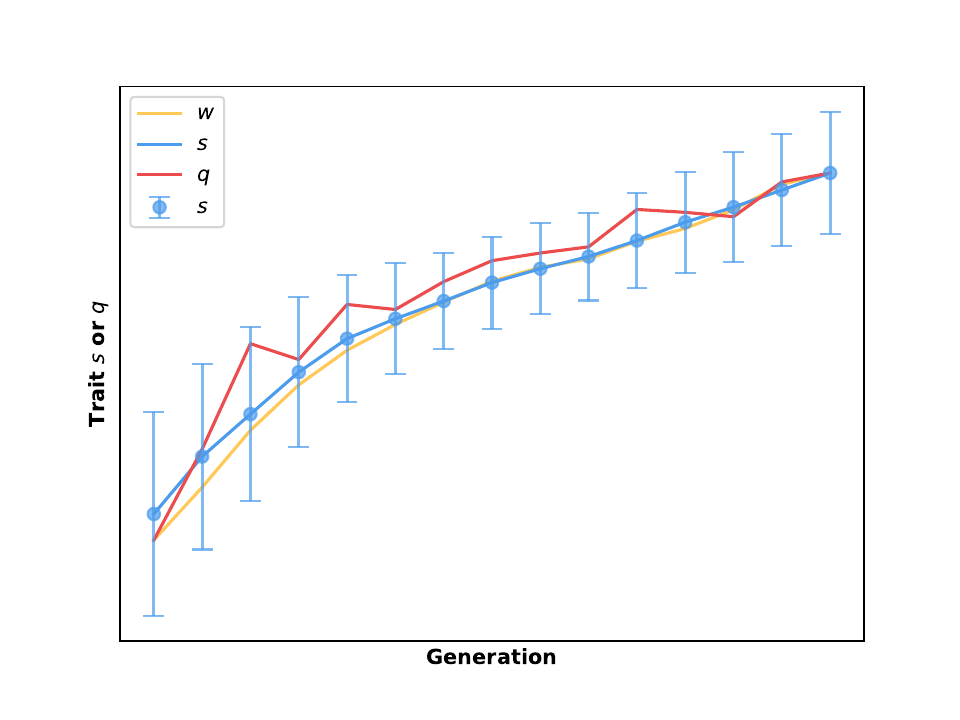}
            \caption*{\scriptsize Natural Selection}
            \label{fig:plot-ns-1}
        \end{subfigure}
        \begin{subfigure}[t]{0.23\textwidth}
            \centering
            \includegraphics[width=\textwidth]{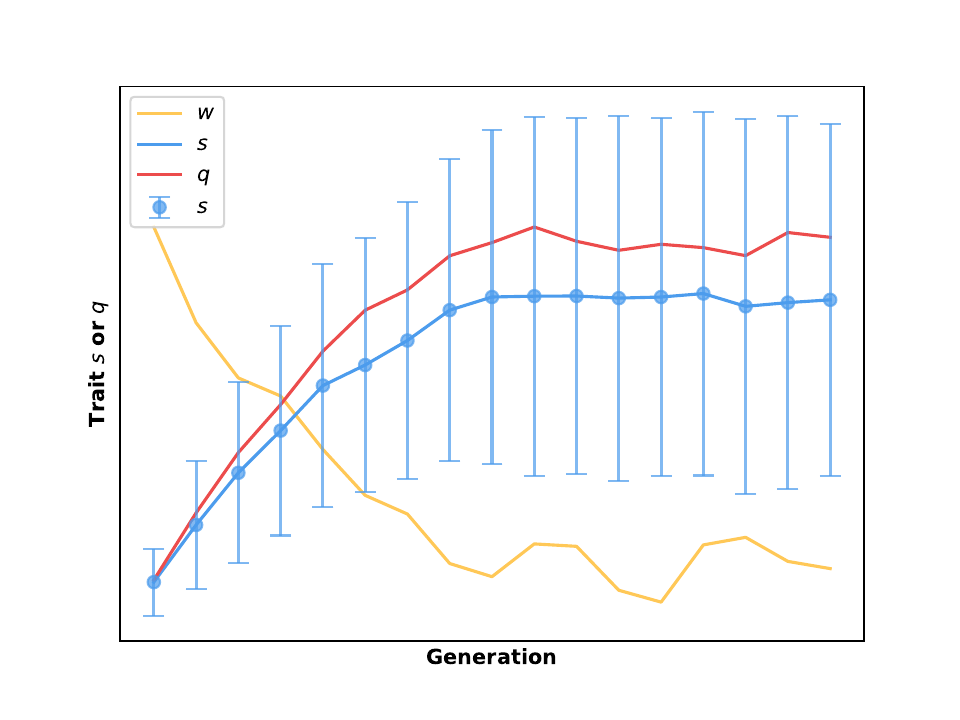}
            \caption*{\scriptsize Overwhelmed By Mutation}
            \label{fig:plot-ns-inverse-1}
        \end{subfigure}
        \begin{subfigure}[t]{0.23\textwidth}
            \centering
            \includegraphics[width=\textwidth]{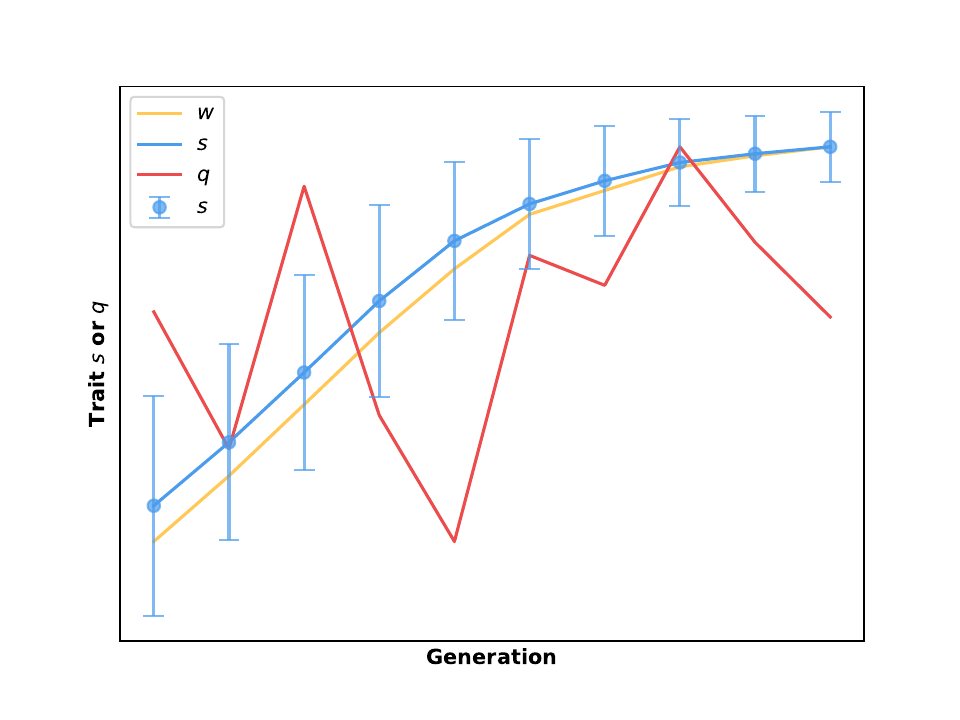}
            \caption*{\scriptsize Exceeds QSL}
            \label{fig:plot-ns-qsl-1}
        \end{subfigure}
        \begin{subfigure}[t]{0.23\textwidth}
            \centering
            \includegraphics[width=\textwidth]{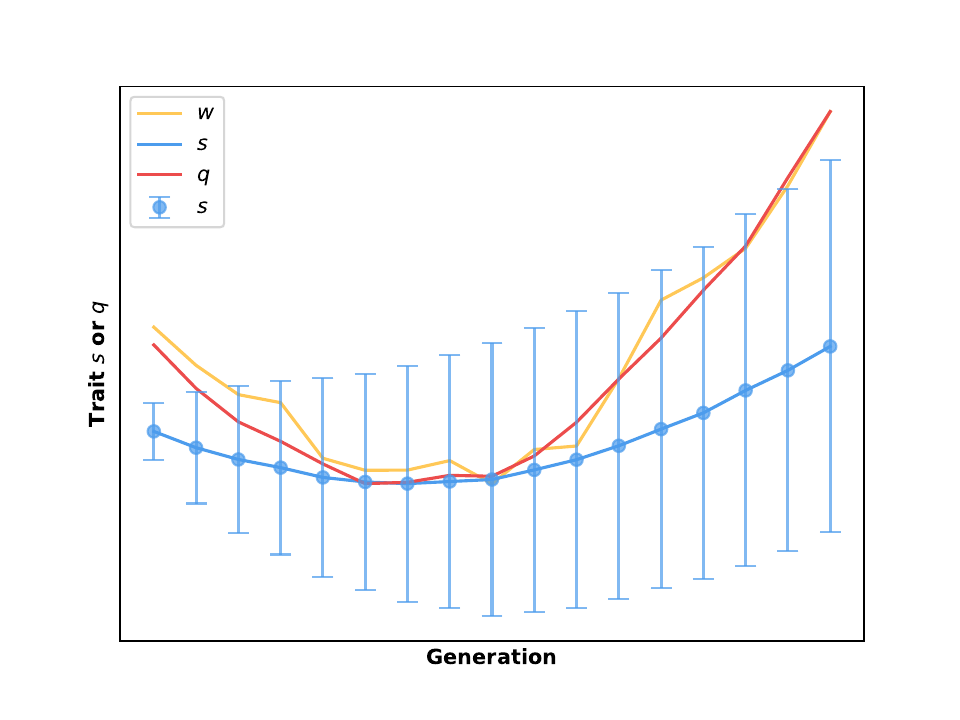}
            \caption*{\scriptsize Rebound Selection}
            \label{fig:plot-ns-rebound-1}
        \end{subfigure}
        \caption*{\scriptsize Blue = \emph{Structural System} \quad  Red = \emph{Qualitative System} \quad  Yellow = \emph{Fitness}}
        \label{fig:ns-grid-1}
    \end{figure}

    \textbf{Implications.} QSL imposes theoretical and empirical constraints on viable structure–quality mappings in any theory of consciousness compatible with natural selection.
    The framework generalizes to any trait pairs mediated by imperfect mappings; the results are not limited to qualia.
    Placement within a broader philosophical context can be found in the companion paper \emph{Structure \& Quality}.
    \restoregeometry 
    \clearpage
    \pagenumbering{arabic}  
    \newpage

    \maketitle
    \begin{abstract}
        This paper explores foundational questions about the relationship of qualia to natural selection.
        The primary result is a derivation of specific formal conditions under which structural
        systems subject to natural selection can convey consistent effects in an associated qualitative domain, placing
        theoretical and empirical constraints on theories of consciousness.
        In order to achieve this result, information-theoretic measures are developed to quantify the mutual
        determinability between structure and quality, quantifying fidelity between the two domains.
        The fidelities represented by that space are then incorporated into the Price Equation
        to yield key bounds on the transmission of selective effects between domains.
        Finally, transmission of higher-order structures between domains is explored.
        Placement within a broader philosophical context can be found in the companion paper \emph{Structure \& Quality}.\footnote{\cite{williams2025structure}}
    \end{abstract}

    It remains an open question what role our qualitative experience plays in our own function as organisms, and
    by extension, what role those qualia might play in the principles of self-organization—including natural selection—
    that gave and give rise to sentient beings.\footnote{\cite{chalmers1995}, \cite{hoffman2015}}
    The essential difficulty is this: Even if we take it as given that the world is somehow possessed of an essentially
    qualitative character, why should that qualitative character manifest in a way that is functionally relevant for
    higher organisms?\footnote{\cite{Lacalli2021}, \cite{Lacalli2024}}

    Even if we assume it to be a rather indirect relationship, it is hard to doubt that our perceptions recapitulate some
    analogous structure within the world.
    If we further consider the apparent role that our planning, deliberations, and desires have on our own behavior,
    we find ourselves situated in a place where qualitative aspects of our experience both cause and are caused by
    (in some broad sense of the notion ``cause'') structural features of the world.
    This situation demands an explanation.

    We could easily imagine a world which was essentially qualitative in character, but in which the structure of its
    qualities bore no relation to the specific concerns of higher organisms—enormous orchestras of
    the stuff of the world, each organized to serve the function of its own autopoiesis.
    In this easily-imaginable scenario, every electron, quark, or other quantum of activity might be the manifestation
    of some qualitative property (the color red or the smell of coffee).

    But we would find no coherence between these atoms of quality, only a radio static of incoherent sensations,
    appearing and then gone again as quickly as they came.
    No permanence, no coherence, no function.

    This paper explores fundamental questions related to this situation:
    \begin{enumerate}
    \item Can we quantify relationships between structure and quality?
    \item Do any of these relationships have the property that selection on a purely structural basis would produce a bias
    towards certain qualitative states?
    \item How can higher-order structures convey from the structural domain into the qualitative domain and vice versa?
    \end{enumerate}

    \section{Background}\label{sec:background}

    In this first section, we will provide some basic philosophical background about what we mean by the terms
    \emph{structure} and \emph{quality}.
    Further discussion can be found in the companion article \emph{Structure \& Quality}.

    \subsection{What Are Structure \& Quality?}\label{subsec:what-are}

    We will use the term \emph{structure}, largely interchangeably with the terms \emph{functional}, \emph{relational},
    and \emph{quantitative}.
    Collectively these all refer to properties that have to do with the arrangement, measure, or relation of things,
    but disregard their intrinsic nature.
    For instance, Newton's equations tell us approximately how matter will move under appropriate macroscopic
    circumstances.
    The movements, measurements, distances, speeds, etc.\ which can be characterized by abstract mathematical or
    computational devices will fall under this structural category.

    Qualitative properties, on the other hand, have some ineffable factor, which can only
    become manifest in things which exhibit those qualitative properties in their own existence.
    To us, these are largely known as sensations and are sometimes referred to as qualia.
    These are characteristics like the experienced scent of freshly baked bread or the greenness of a tree.

    For the purposes of this paper, we will suppose that there is no way that one can derive qualitative properties
    by principle from structural properties alone.
    Instead we assume that we can only specify the relationships between structure and quality, without supposing any
    necessary connection between them.

    This is exactly the condition we find ourselves in regarding fundamental questions of ontology under normal
    circumstances of scientific investigation.
    We continually produce more powerful descriptions of nature, but none that hold out of necessity from first
    principles.
    So we have ever more exact specifications of how an electron behaves, but at bottom no indication why there need be
    electrons at all.
    (And if there were a reason discovered why there were electrons, \emph{that} explanation would then be in need of
    an explanation of its own.)

    In the same way, we will give descriptions of structure, and we will give descriptions of how qualities
    correlate with that structure, and even descriptions of how the correlations might have correlations.
    But ultimately, we have taken it for granted that there are qualities there to describe—no further
    explanation feigned.

    We will aim to show persuasively that no such derivation is necessary to establish the functional connection to
    qualitative properties that is necessitated by natural selection and other principles of self-organization.

    \section{Informatics}\label{sec:informatics}

    We take any given model of a structural-qualitative relationship to hold between a structural domain $S$ and a
    qualitative domain $Q$.
    We make no presumption about the structure of these domains except that their states $s \in S$ and $q \in Q$
    are related by some mapping $\phi : S \to Q$.

    In this setting, we can analyze models according to their informational properties, especially in correlations
    between the domains dictated by the properties of the relation $\phi(s) = q$.
    We can leverage those in conjunction with existing
    evolutionary theory to derive theoretical results about \emph{Q-S} systems generally.

    \subsection{Q-S Space}\label{subsec:qs-space}

    \emph{Q-S Space} is a 2-dimensional representation of the amount of information the structural and qualitative
    spaces determine about each other.
    Any system can be represented as a pair $(\hat{q},\hat{s})$.

    The first coordinate $\hat{q}$ is the level of certainty in the structural state given a known qualitative state,
    the second $\hat{s}$ is the certainty in the qualitative state given a known structural state.
    1 implies perfect certainty, while 0 indicates no correlation.

    Each coordinate can be calculated as follows:
    \begin{equation} \label{eq:sq-hat}
        \hat{q} = \eta_{s \mid q}, \quad \hat{s} = \eta_{q \mid s},
    \end{equation}
    where $\eta$ is the correlation ratio.

    The correlation ratio $\eta$ is defined as:
    \begin{equation}\label{eq:c-continuous}
    \eta_{a \mid b} = \sqrt{\frac{\Var(\expectation[a \mid b])}{\Var(a)}}.
    \end{equation}
    Intuitively, $\eta_{a \mid b}$ is the amount of uncertainty in $a$ explained by $b$, normalized in the range $[0,1]$.
    Further discussion on \emph{Q–S systems} can be found in the companion paper \emph{Structure \& Quality}.

    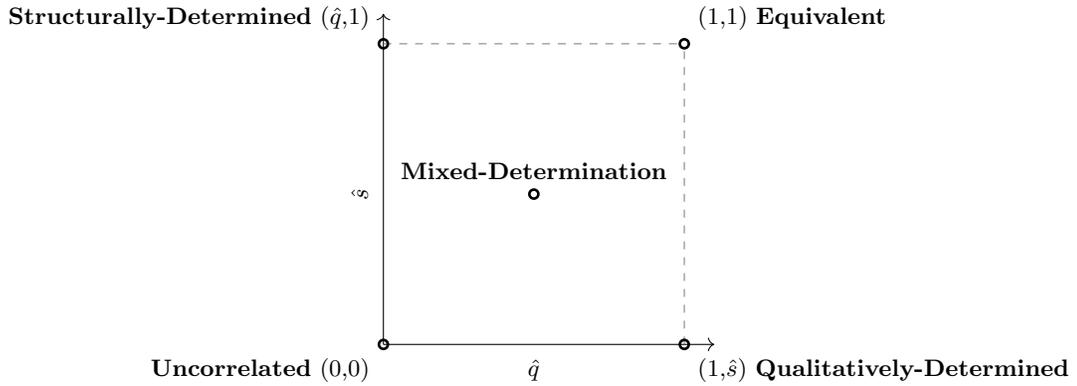
\begin{figure}[htbp]
        \centering
        \begin{tikzpicture}[scale=4, font=\small]

            \draw[->] (0,0) -- (1.1,0);
            \draw[->] (0,0) -- (0,1.1);

            \node[rotate=0, anchor=north, font=\small] at (0.5, -0.02) {$\hat{q}$};
            \node[rotate=90, anchor=south, font=\small] at (-0.02, 0.5) {$\hat{s}$};

            \draw[dashed, gray] (1,0) -- (1,1);
            \draw[dashed, gray] (0,1) -- (1,1);

            \node[circle, draw, inner sep=1.2pt, line width=1pt, label=below left:{\textbf{Uncorrelated} (0,0)}] at (0,0) {};
            \node[circle, draw, inner sep=1.2pt, line width=1pt, label=above left:{\textbf{Structurally-Determined} ($\hat{q}$,1)}] at (0,1) {};
            \node[circle, draw, inner sep=1.2pt, line width=1pt, label=below right:{(1,$\hat{s}$) \textbf{Qualitatively-Determined}}] at (1,0) {};
            \node[circle, draw, inner sep=1.2pt, line width=1pt, label=above right:{(1,1) \textbf{Equivalent}}] at (1,1) {};
            \node[circle, draw, inner sep=1.2pt, line width=1pt, label=above:{\textbf{Mixed-Determination}}] at (0.5,0.5) {};

        \end{tikzpicture}
        \caption{Five model types represented as $(\hat{q},\hat{s})$ pairs in \emph{Q-S space}.
        Points that lie on the dashed lines (where either or both coordinates are 1) are referred to as \emph{fully-determinate}.
        Either their quality or their structure (or both) is sufficient to describe the entire system, so long as the appropriate
        derivation of the other is known.}
        \label{fig:entropy-map}
    \end{figure}

    \begin{figure}[htbp]
        \centering
        \includegraphics[width=0.85\textwidth]{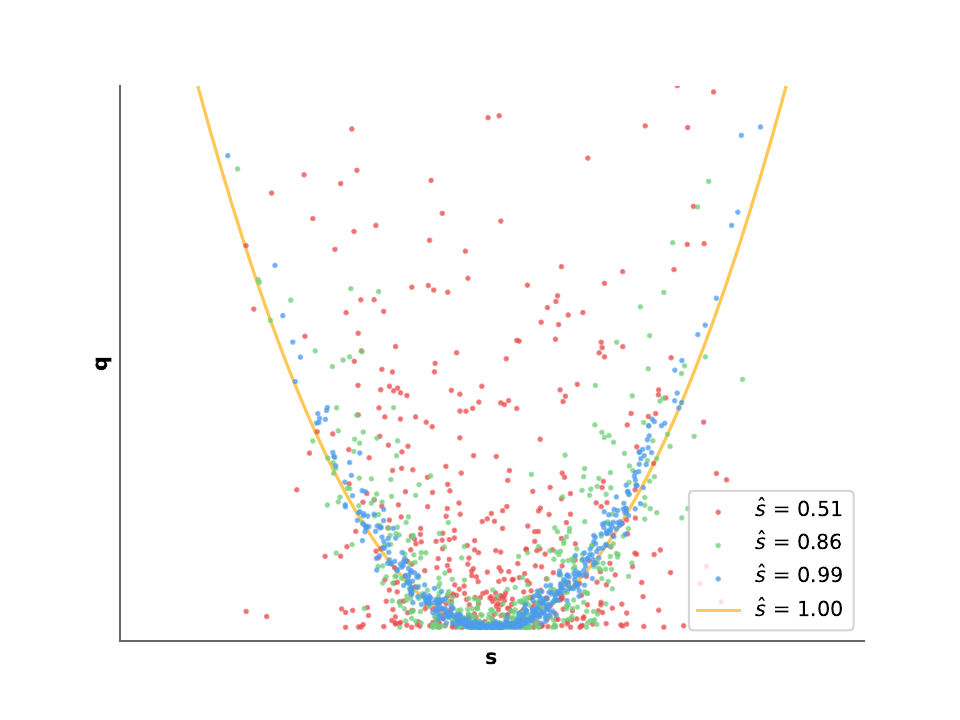}
        \caption{
            Scatterplot illustrating the fidelity between $q$ and $s$ for different values of $\hat{s}$.
            This data was derived from simulations which modeled $\phi(s)$ as approximately quadratic, with the result that
            a perfectly correlated $q$ would follow a parabolic curve.
            Noise was injected to model different levels of correlation between $q$ and $s$.
        }
        \label{fig:q_s_fidelity}
    \end{figure}

    \section{Natural Selection}\label{sec:natural-selection}

    Turning toward natural systems, we find that applicability of our system includes exploring how selection within one
    domain can be conveyed to the other, and deriving potential constraints on the state of our system in \emph{Q-S space}
    via known mathematical properties of natural selection.

    In particular, it is known that noise can overwhelm the power of natural selection via transmission effects.
    Usually this noise in biological systems is considered in the forms of mutation, recombination, and drift.
    But in our picture, the translation into the $Q$ domain itself is another source of noise—not in the state of
    the structural system $S$, but in its appearance in the qualitative system via the relation $\phi(s)$.

    We can show:
    \begin{enumerate}
    \item That any system which produces a dependence of quality on structure could ``indirectly'' select for a favored set of qualitative states; and
    \item That we can derive specific conditions on the fidelity between the domains for when selection can be conveyed effectively from one domain into the other.
    \end{enumerate}
    \subsection{Qualitative Selection}\label{subsec:intrinsic-informatics}

    Imagine we have some fitness function $f : S \to \mathbb{R}$ which favors some subset of structural states
    $S^* \subset S$.
    Natural selection acts to modify the distribution $P^*(s)$ over $S$ such that
    \begin{equation}\label{eq:selection}
    P^*(s) = \begin{cases}
                 \gg 0 & \text{if } s \in S^*\\
                 \approx 0 & \text{if } s \notin S^*.
    \end{cases}
    \end{equation}

    In cases where the structure holds influence over a system's qualitative properties, i.e.\ where $(\hat{q},\hat{s} > 0)$ in
    \emph{Q-S} terms, that favored subset $S^*$ will also pick out a corresponding subset $Q^* \subset Q$.
    We can marginalize out $P(q)$ for some $q \in Q$ as
    \begin{equation}\label{eq:q-selection}
    P(q) = \sum_{s \in S} P(q \mid s)\,P^*(s).
    \end{equation}
    Since $P^*(s) \approx 0$ for $s \notin S^*$,
    \begin{equation}\label{eq:q-selection-derived}
    P(q) \approx \sum_{s \in S^*} P(q \mid s)\,P^*(s).
    \end{equation}
    And therefore a set of qualitative states $Q^*$, correlated with the structurally-favored states $S^*$ will become
    overrepresented in the post-selection distribution, even if they are not necessarily being directly acted upon by
    any selective pressure.
    Any conclusions made here that apply from the structural to the qualitative domain should also apply equally in
    the opposite direction for systems where $(\hat{q} > 0,\hat{s})$.

    \subsection{The Price Equation}\label{subsec:price-equation}

    The Price Equation\footnote{\cite{price1970}} defines a framework for analyzing systems under natural selection.
    \begin{equation}\label{eq:price-equation}
    \Delta \bar{z} =
    \underbracket{\Cov(w_i, z_i)}_{\text{selection}} +
    \underbracket{\expectation[w_i \Delta z_i]}_{\text{transmission}}
        \text{ where } \Delta z_i = \bar{z_i}' - z_i
    \end{equation}

    \begin{itemize}
    \item $\Delta \bar{z}$ represents the average rate of change in a trait $z$ across generations.
    \item $\Cov(w_i, z_i)$ is the term that represents the effect of selection; it indicates the shared variation
    between $z_i$ (the trait $z$ in an individual $i$) and $w_i$ (the fitness of an individual $i$
    represented as a number of offspring).
    \item $\expectation[w_i \Delta z_i]$ represents how the trait $z$ changes between parents and offspring, the expectation of
    the average fitness (expressed above) scaled by $\Delta z_i$, the change in trait $z$ for an individual $i$
    where $\bar{z_i}'$ is the average value of $z$ among the offspring.
    \end{itemize}

    We can see that when
    \begin{equation}\label{eq:overwhelmed}
    \abs{\frac{\expectation[w_i \Delta z_i]}{\Cov(w_i, z_i)}} \gg 1
    \end{equation}
    the selection term is overwhelmed by other transmission effects, and the selected result is no longer stable.

    \subsection{Correlations Between Domains}\label{subsec:q-minima}

    Recall that the mapping \(\phi: S \to Q\) is characterized by a fidelity \(\hat{s} = \eta_{q \mid s}\), defined in (\ref{eq:sq-hat})
    with \(\hat{s}=1\) representing perfect determination of the qualitative state given the structural state.
    If we assume that $q$ can be estimated via a mapping $\bar{\phi} : S \to Q$, then
    \begin{equation}\label{eq:s-hat-new}
    \hat{s} = \eta_{q \mid s} = \sqrt{\frac{\Var(\expectation[q \mid s])}{\Var(q)}}.
    \end{equation}
    We can define $\bar{\phi}$ as
    \begin{equation}\label{eq:phi-def}
    \phi(s) = \bar{\phi}(s) + \epsilon,
    \end{equation}
    where $\bar{\phi}(s)$ accounts for the part of $q$ in correlation with $s$ and $\epsilon$ is the remaining variation.
    Substituting $\bar{\phi}(s)$ for $\expectation[q \mid s]$ we have
    \begin{equation}\label{eq:phi-bar}
    \hat{s} = \sqrt{\frac{\Var(\bar{\phi}(s))}{\Var(q)}.}
    \end{equation}

    We can substitute a structural trait $s \in S$ or qualitative trait $q \in Q$ into the Price Equation to yield:
    \begin{equation}\label{eq:price-subst-s}
    \Delta \bar{s} = \Cov(w_i, s_i) + \expectation[w_i\,\Delta s_i]
    \end{equation}
    \begin{equation}\label{eq:price-subst-q}
    \Delta \bar{q} = \Cov(w_i,q_i) + \expectation[w_i\,\Delta q_i].
    \end{equation}

    From\ \eqref{eq:phi-def} and\ \eqref{eq:price-subst-q} we can see that
    \begin{equation}\label{eq:cov-wq}
    \Cov(w_i,q_i) = \Cov(w_i,\phi(s)) = \Cov(w_i,\bar{\phi}(s)) + \Cov(w_i,\epsilon).
    \end{equation}
    Since we can assume $w$ is uncorrelated with $\epsilon$, we have
    \begin{equation}\label{eq:cov-wq-approx}
    \Cov(w_i,q_i) \approx \Cov(w_i,\bar{\phi}(s_i)),
    \end{equation}
    and applying the Cauchy-Schwartz inequality, we can say
    \begin{equation}\label{eq:cauchy}
    \abs{\Cov(w_i,\bar{\phi}(s_i))} \leq \sqrt {\Var(w_i) \Var(\bar{\phi}(s_i))}.
    \end{equation}

    If we note that $\Var(\bar{\phi}(s_i)) = \hat{s}^2 \Var(q_i)$ we obtain
    \begin{equation}\label{eq:cauchy-subst}
    \abs{\Cov(w_i,q_i)} \lesssim \hat{s} \, \sqrt {\Var(w_i) \Var(q_i)},
    \end{equation}
    meaning our qualitative selection term is proportional to (or in other words ``dampened'' by) the fidelity $\hat{s}$.

    Dividing both sides by $\sqrt {\Var(w_i) \Var(q_i)}$ we arrive at the bound
    \begin{equation}\label{eq:wq-bound}
    \abs{\rho_{w,q}} \lesssim \hat{s}
    \end{equation}
    where $\rho_{w,q}$ is the Pearson correlation coefficient
    \begin{equation}\label{eq:pearson}
    \rho_{a,b} = \frac{\Cov(a, b)}{\sqrt{\Var(a) \Var(b)}}.
    \end{equation}

    Finally, assuming $w$ is only correlated with $q$ through $s$, we obtain the following via the data processing inequality:
    \begin{equation}\label{eq:result}
    \abs{\rho_{w,q}} \lesssim \hat{s} \cdot \abs{\rho_{w,s}},
    \end{equation}
    and therefore
    \begin{equation}\label{eq:final-result}
    \abs{\frac{\rho_{w,q}}{\rho_{w,s}}} \lesssim \hat{s}.
    \end{equation}
    This is a potentially significant result which shows that the correlation between $w$ and $q$ is bounded by the correlation
    between $w$ and $s$, scaled by $\hat{s}$.
    It has potential for empirical applications in evaluating proposed mappings $\phi$ between the structural and
    qualitative domains.

    \subsection{Approximating Transmission}\label{subsec:approximating-transmission}

    We'll first take a closer look at the transmission term from (\ref{eq:price-equation}),
    \begin{equation}\label{eq:transmission}
    \expectation[w_i \Delta z_i] \text{ where } \Delta z_i = \bar{z_i}' - z_i.
    \end{equation}
    Recall that $\Delta z_i$ represents non-selectional variation from sources like mutation, recombination, and genetic
    drift.
    For the purposes of our discussion, we will assume that $\Delta z_i$ is uncorrelated with $z$ and with $w$.

    In the context of the qualitative Price Equation (\ref{eq:price-subst-q}), we can state the transmission term as
    $\expectation[w_i \Delta q_i]$, where $q_i$ may be approximated
    \begin{equation}\label{eq:q-transmission}
    \Delta q_i \approx \bar{\phi}(s_i') - \bar{\phi}(s_i).
    \end{equation}
    In some cases it may be useful to approximate $\Delta q_i$ by way of the derivative $\bar{\phi}'$ as opposed to
    $\bar{\phi}$ directly.
    If we assume $\Delta s_i$ is small and we can approximate $\bar{\phi}(s)$ as locally linear we obtain
    \begin{equation}\label{eq:delta-q-approx}
    \Delta q_i \approx \bar{\phi}'(s_i) \, \Delta s_i.
    \end{equation}

    \subsection{The Qualitative Selection Limit}\label{subsec:qql}

    Now combining the results from the previous sections, we can substitute back into (\ref{eq:price-subst-q}) to derive
    \begin{equation}\label{eq:q-price}
    \Delta \bar{q} \lesssim \hat{s} \, \sqrt {\Var(w_i) \Var(q_i)} + \expectation[w_i \, \Delta q_i].
    \end{equation}
    Moving further, we can substitute back into (\ref{eq:overwhelmed}) to obtain the following
    \emph{qualitative selection limit} (QSL),
    \begin{equation}\label{eq:qdc}
    \abs{\frac{\expectation[w_i \, \Delta q_i]}{\sqrt {\Var(w_i) \Var(q_i)}}} \gg \hat{s}.
    \end{equation}

    When\ \eqref{eq:qdc} obtains, the selective pressure on the structural domain is no longer conveyed faithfully into
    the qualitative domain.
    This salient result places a bound on the conditions that a selective pressure from the structural domain could be
    conveyed into the qualitative domain, without being overwhelmed by transmission effects in the qualitative domain.

    The key takeaway is that selective pressures which might favor specific qualities in the structural domain will only
    translate into the qualitative domain under certain constraints, before the selective signal is dampened out.

    \subsection{Structural Stability}\label{subsec:structural-stability}

    Intuitively, it seems additional constraints may come into play centered on the
    stability of the mapping $\phi : S \to Q$.
    A mapping is considered structurally-stable if its mapped features do not change under small perturbations.\footnote{For discussion on
    structural stability in biological contexts see \cite{Thom1975}.}
    In fact, some of the approximations in the previous section are only applicable under these conditions of stability.

    This can be expressed as a Lipschitz condition on $\phi$ that there exists a constant $L$ such that for all
    $s_1, s_2 \in S$,
    \begin{equation}\label{eq:lipschitz}
    \lVert \phi(s_1) - \phi(s_2) \rVert \ \leqq \ L \, \lVert s_1 - s_2 \rVert.
    \end{equation}
    This condition ensures that small changes in $S$ lead to proportionally small changes in $Q$.

    In practice, this condition may only need to hold within some subset of the state space $S' \subset S$ which is
    under selective pressure.
    The requirement for structurally stable mappings between $S$ and $Q$ may also be a key condition that places
    rather strict bounds on the potential mappings that need to be considered.

    This idea can be generalized from using the additive difference as a distance metric to employing a more specific
    distance defined on the structural and qualitative spaces $S$ and $Q$
    \begin{equation}\label{eq:lipschitz-adapted}
    \mathbb{D}_Q\big[ \phi(s_1) \, \lVert \, \phi(s_2) \big] \ \leqq \ L \, \mathbb{D}_S\big[ s_1 \, \lVert \, s_2 \big].
    \end{equation}
    where $\mathbb{D}_S$ and $\mathbb{D}_Q$ are distances defined on each space, respectively.

    \section{Higher-Order Constructs}\label{sec:higher-order-constructs}

    Though we've established a basis for analyzing \emph{Q-S} systems, there are still significant hurdles that have not
    been discussed.
    In the actual case of biological entities we observe, both the organism\footnote{\cite{Thompson1917}, \cite{Niklas2016}}
    and the manifest experience\footnote{\cite{marr1982}, \cite{dennett1991}, \cite{williams2022}} are
    highly-structured, many levels above any fundamental constituents, often with nontrivial connections to the
    fundamental constituents.\footnote{\cite{Rosen1991}, \cite{Rosen2000}}

    So while the \emph{Q-S} correlations we've discussed are critical conditions for the conveyance of selectional
    pressures between systems, it remains to be demonstrated under what circumstances those correlations will cascade
    to higher-order structures.
    A cursory investigation in this direction follows.

    \subsection{Higher Levels as Informative Priors}\label{subsec:informative-priors}

    A higher level of analysis of some state can be naturally seen as inducing an informative prior on its base state.
    If we denote the state of a system at level $\ell$ as $s_\ell$, we may state the relation between levels as a relation
    $\omega$.
    \begin{equation}\label{eq:omega}
    \omega(s_{\ell}) = s_{\ell+1}.
    \end{equation}
    Now we note that, each $s_{\ell+1} \in S_{\ell+1}$ maps to potentially more than one $s_\ell \in S_\ell$, a feature
    important for coarse-graining and dimension reduction.

    Because of this property, knowing $s_{\ell+1}$ allows us to pick out some subset
    of possible states
    \begin{equation}\label{eq:sl-possible-states}
    S_\ell^* = \{ s_\ell : \omega(s_{\ell}) = s_{\ell+1} \}
    \end{equation}
    such that the marginal distribution satisfies
    \begin{equation}\label{eq:sl-favored-states}
    P(s_{\ell+1}) = \sum_{s_\ell \in S_\ell^*} P(s_\ell),
    \end{equation}
    and conditioning on $s_{\ell+1}$ defines an informative prior over $s_\ell$,
    \begin{equation}\label{eq:sl-prior}
    P(s_\ell \mid s_{\ell+1}) < P(s_\ell).
    \end{equation}
    We may use the knowledge of $s_{\ell+1}$ to refine our expectation about the lower-level state $s_{\ell}$.
    See~\hyperref[sec:higher-examples]{Appendix B} for examples.

    \subsection{Vector Spaces}\label{subsec:vector-spaces}

    Let's now take $S_0 = \mathbb{R}^2$ and $S_1 = \mathbb{R}$.
    This is a clear example of dimensional reduction from two dimensions to one.
    We can utilize the properties of vector spaces to draw conclusions about the potential relationships of these spaces.
    Each $s_0$ will be represented as a pair $(x,y)$.

    In our first example, our mapping $\omega$ will simply discard the $y$ coordinate:
    \begin{equation}\label{eq:discard-y}
    \omega(x, y) = x.
    \end{equation}
    If we consider the fidelity of the mapping we can say $\hat{\omega}_x = 1$ ($\omega$ is perfectly informative about
    $x$) while $\hat{\omega}_y = 0$ ($\omega$ is uninformative informative about $y$).

    Now if we generalize $\omega$ as a linear combination of $x$ and $y$
    \begin{equation}\label{eq:linear-combination}
    \omega(x, y) = ax + by
    \end{equation}
    Let us suppose \( x \) and \( y \) are jointly distributed with some known covariance structure.
    Then the fidelity of \( \omega \) with respect to $x$ can be evaluated using the correlation ratio:
    \begin{equation}\label{eq:hat-omega-x}
        \hat{x}_\omega = \eta_{\omega \mid x}
            = \sqrt{\frac{\Var(\expectation[\omega \mid x])}{\Var(\omega)}}
            = \sqrt{\frac{\Var(ax + b \expectation[y \mid x])}{\Var(\omega)}},
    \end{equation}
    where
    \begin{equation}\label{eq:var-omega}
    \Var(\omega) = a^2 \Var(x) + b^2 \Var(y) + 2ab \Cov(x, y).
    \end{equation}
    This quantity describes how much of the variance in \( \omega \) can be explained by conditioning on
    \( x \).
    Similar methods may be used for $y$.

    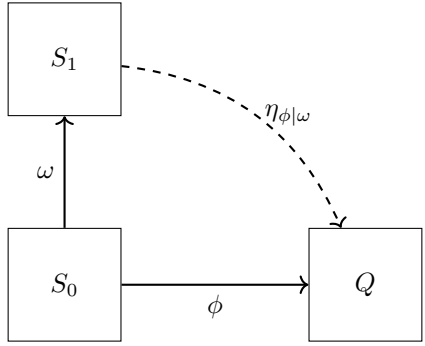
\begin{figure}[htbp]
        \centering
        \begin{tikzpicture}[
            node distance=6cm,
            boxnode/.style={
                draw, rectangle, minimum width=1.5cm, minimum height=1.5cm, align=center, font=\normalsize
            }
        ]
            \node[boxnode] (S0) at (0,0) {\( S_0 \)};
            \node[boxnode] (S1) at (0,3) {\( S_1 \)};
            \node[boxnode] (Q)  at (4,0) {\( Q \)};

            \draw[->, thick] (S0) -- node[left] {\( \omega \)} (S1);
            \draw[->, thick] (S0) -- node[below] {\( \phi \)} (Q);
            \draw[->, dashed, thick, bend left=30] (S1) to node[right] {$\eta_{\phi \mid \omega}$} (Q);
        \end{tikzpicture}
        \caption{$S_0 = \mathbb{R}^2$ is mapped into $S_1 = \mathbb{R}$ and $Q = \mathbb{R}$ via the transforations $\omega$ and $\phi$ separately, but
        still share a fidelity $\eta_{\phi \mid \omega}$.
        In the linear case, when $\eta_{\phi \mid \omega} \approx 1$, the mappings are close to linearly-dependent and can approximate each other scaling by a constant $\lambda$.}
        \label{fig:omega-diagram}
    \end{figure}

    \subsection{Fidelities}\label{subsec:fildelity-transmission}

    Let's now examine the case of a system where we have separate spaces $S_0 = \mathbb{R}^2$,
    $S_1 = \mathbb{R}$, and $Q = \mathbb{R}$ with
    \begin{equation}\label{eq:omega-1}
        \omega : S_0 \to S_1
    \end{equation}
    and
    \begin{equation}\label{eq:phi-1}
        \phi : S_0 \to Q.
    \end{equation}

    We can quantify their fidelity $\eta_{\phi \mid \omega}$ as the correlation ratio from $\phi$ to $\omega$:
    \begin{equation}\label{eq:eta-omega-given-phi}
    \eta_{\phi \mid \omega}
    = \sqrt{ \frac{ \Var\left( \mathbb{E}[\phi \mid \omega] \right) }{ \Var(\omega) } }.
    \end{equation}
    In the linear case, we can say:
    \begin{equation}\label{eq:eta-omega-given-phi-linear}
    \eta_{\phi \mid \omega} = \frac{ \Cov(\omega, \phi) }{ \sqrt{ \Var(\omega) \Var(\phi) } }
    = |\rho_{\omega, \phi}|,
    \end{equation}
    where $\rho_{\omega, \phi}$ is the correlation coefficient between $\omega$ and $\phi$.

    In a simple case where both $\omega$ and $\phi$ are linear combinations of $x$ and $y$:
    \begin{equation}\label{eq:linear-combination-omega}
    \omega(x, y) = ax + by,
    \end{equation}
    \begin{equation}\label{eq:linear-combination-phi}
    \phi(x, y) = cx + dy.
    \end{equation}
    With some basic linear algebra, we can see that $\phi$ can be derived from $\omega$ (and vice versa) when $[a \ b]$ and $[c \ d]$
    are linearly dependent:
    \begin{equation}\label{eq:linear-dependence}
    \omega(x, y) = \lambda \, \phi(x,y) \text{ when } \lambda = \frac{a}{c} = \frac{b}{d}.
    \end{equation}
    This case reflects perfect fidelity $\eta_{\phi \mid \omega} = 1$, and fidelity drops as the functions approach orthogonality.

    \begin{figure}[htbp]
        \centering
        \includegraphics[width=.7\textwidth]{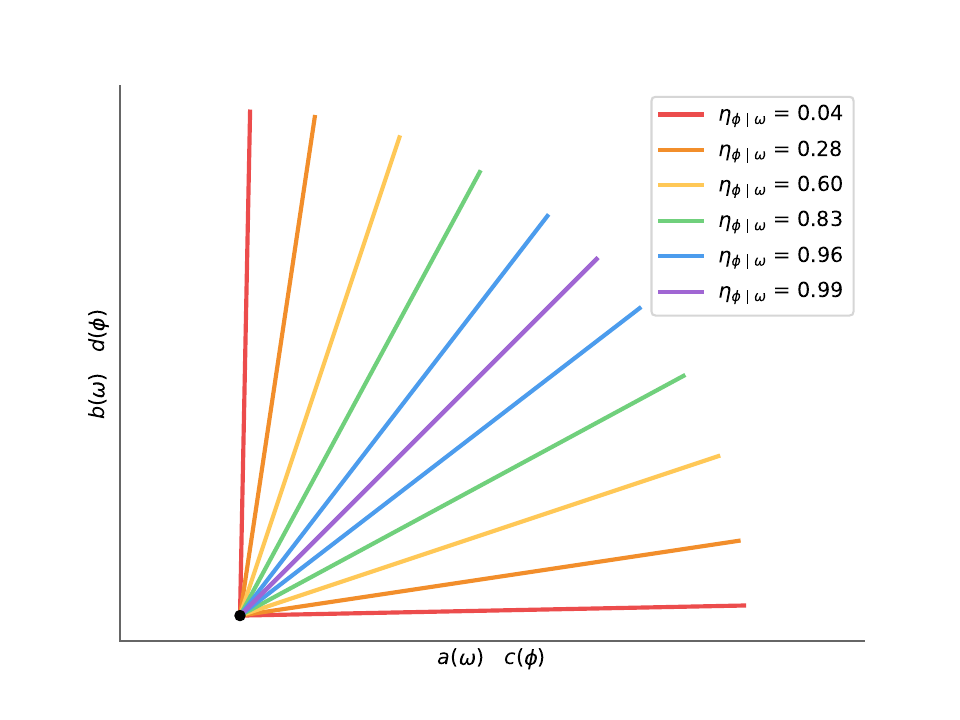}
        \caption{$(a,b)$ and $(c,d)$ under a range of fidelities $\eta_{\phi \mid \omega}$.
        This figure plots the parameters of $\omega$ ($a$ vs. $b$) and of $\phi$ ($c$ vs.\ $d$), demonstrating that the closer
        to linear dependence the parameters are, the higher their fidelity.}
        \label{fig:omega-phi}
    \end{figure}

    \begin{figure}[htbp]
        \centering
        \includegraphics[width=.7\textwidth]{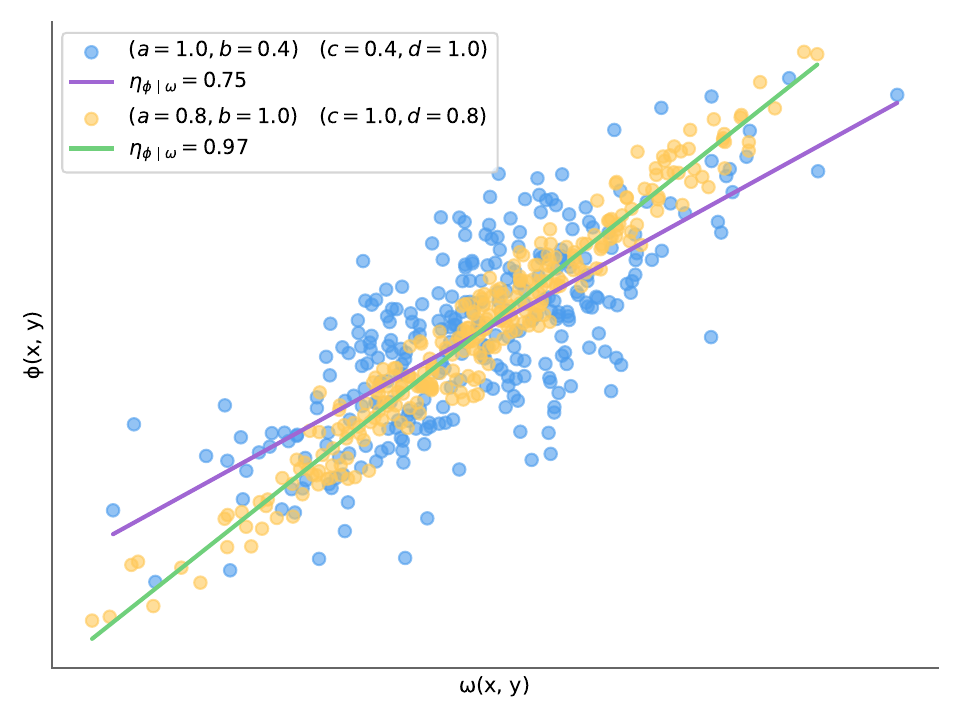}
        \caption{$\omega$ vs.\ $\phi$. Two distributions of $\omega$ vs $\phi$ values over a random distribution of $(x,y)$ pairs.
        The higher their fidelity $\eta_{\phi \mid \omega}$, the tighter the distribution around their best fit line.}
        \label{fig:omega-phi-2}
    \end{figure}

    \subsection{Higher-Order Conveyance}\label{subsec:higher-order-conveyance}

    Now we can examine the conditions required for higher-ordered structures to convey from $S_1$ to $Q$
    via $S_0$.

    First we notice that\ \eqref{eq:sl-favored-states} displays a selective affect similar to\ \eqref{eq:q-selection-derived}.
    This will bias the states $s_0$ to the set of states $S^*_0$ if we condition on some known $s_1$.
    We can then examine how that favored set of states is conveyed into the qualitative states via $\phi$.

    When we condition on the set of states $S^*_0$ favored by our known $s_1$, we derive
    \begin{equation}\label{eq:phi-pushforward}
    \expectation[\phi \mid \omega] = \expectation[q \mid s_1] = \sum_{s \in S^*_0} P(s_0 \mid s_1) \  \bar{\phi}(s_0).
    \end{equation}
    Selection over $S_0$ induces a conditional distribution over $Q$ through the mapping $\phi$.

    Now substituting back into~\eqref{eq:eta-omega-given-phi}, we have
    \begin{equation}\label{eq:eta-q-given-s1}
    \eta_{\phi \mid \omega}
    = \sqrt{ \frac{ \Var\left( \expectation[q \mid s_1] \right) }{ \Var(q)} }.
    \end{equation}

    When $\phi$ is highly aligned with $\omega$ via the fidelity $\eta_{\phi \mid \omega} \gg 0$ higher-order properties
    found in $S_1$ will be conveyed into $Q$.
    The correlation ratio $\eta_{\phi \mid \omega}$ functions as a transmission coefficient between levels, similar to
    the fidelity $\hat{s}$.

    \subsection{Generalizations}\label{subsec:generalizations}

    I have noted in the course of this work potential to generalize
    the applicability of these techniques beyond the selection of qualia in particular.

    Following the same logic as~\eqref{eq:result}, we have
    \begin{equation}\label{eq:rho-omega}
    \abs{\rho_{s,\phi}} \lesssim \eta_{\phi \mid \omega} \abs{\rho_{s,\omega}}.
    \end{equation}
    This shows a general course of being able to apply our fidelity measure $\eta$ as a scaling factor to the relevant
    correlation coefficients.
    Under broad conditions this can be done recursively across relations and spanning hierarchies:
    \begin{equation}\label{eq:hierarchy}
    \abs{\rho_{n,0}} \lesssim \eta_{n,n-1} \ldots \,\eta_{2,1} \cdot \abs{\rho_{1,0}}.
    \end{equation}
    Utilizing the same strategies, we derive a general form of the selection limit in~\eqref{eq:qdc}:
    \begin{equation} \label{eq:selection-limit-general}
        \abs{\frac{\expectation[w_i \, \Delta x_i]}{\sqrt {\Var(w_i) \Var(x_i)}}} \gg \eta_{x \mid s}
    \end{equation}
    where, importantly, $s \in S$ is the state of the domain under selection.

    Any instance where selection of a certain trait may indirectly give rise to another trait via some intermediary
    correlation is amenable to the same general methods of inquiry, and
    with subtle changes to the formalism, similar strategies can be used to analyze changing fitness landscapes as
    opposed to different trait spaces.

    \section{Future Directions}\label{sec:future-directions}

    Future investigation of these techniques in the context of specific model systems may be found fruitful.

    Candidate systems include Turing's reaction-diffusion networks,\footnote{Originally \cite{turing1952}, more
    recently explored within the realm of consciousness in \cite{Lacalli2020}, \cite{Lacalli2022a}, and \cite{Lacalli2022b}.}
    Rosen's discussion of the relationship between classical and statistical mechanics,\footnote{\cite{rosen1985}}
    hierarchies within Friston's active-inference systems,\footnote{\cite{friston2010}, \cite{Palacios2020}, \cite{Parr2022}}
    autocatalytic sets and NK systems,\footnote{\cite{kauffman1986autocatalytic}, \cite{kauffman1993origins}}
    and cellular automata.\footnote{\cite{vonneumann1966}}

    \pagebreak

    \bibliographystyle{apalike}
    \bibliography{references}

    \pagebreak

    \appendix

    \section{Appendix A: Computer Simulations}\label{sec:computer-simulations}

    There are computer simulations based on the model in the accompanying materials.

    \subsection{Single-Generation Simulations}\label{subsec:single-gen}

    Below are shown plots displaying the relation between traits and fitness within a single generation dependent on the
    measure $\hat{s}$.

    \begin{figure}[h]
        \centering

        \begin{subfigure}[t]{0.49\textwidth}
            \centering
            \includegraphics[width=\textwidth]{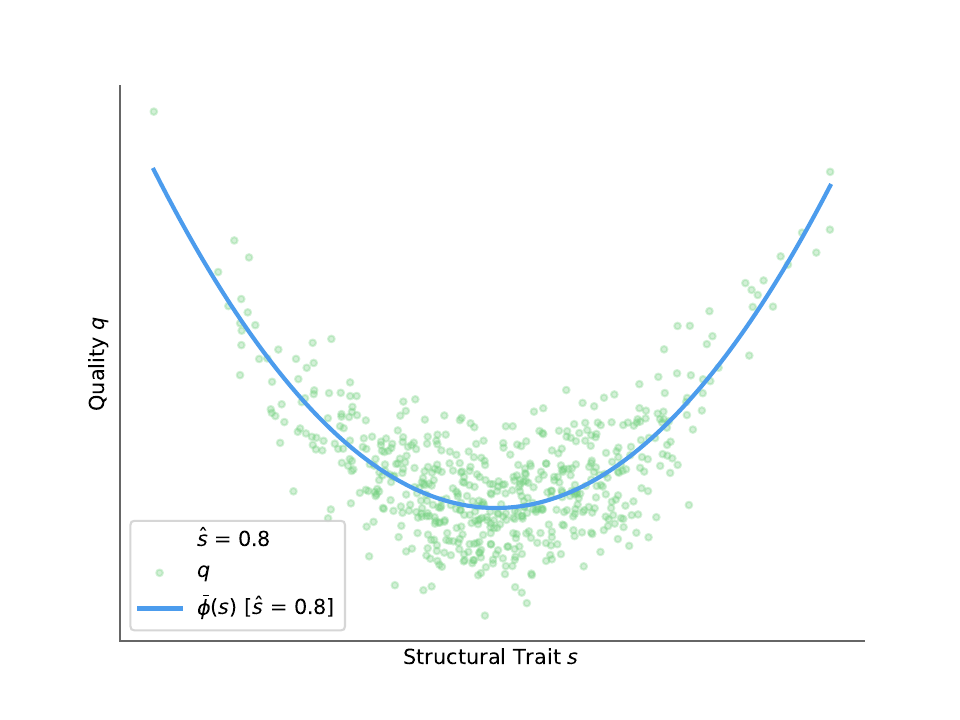}
            \caption{Structural trait $s$ vs qualitative trait $q$}
            \label{fig:plot-sq}
        \end{subfigure}
        \hfill
        \begin{subfigure}[t]{0.49\textwidth}
            \centering
            \includegraphics[width=\textwidth]{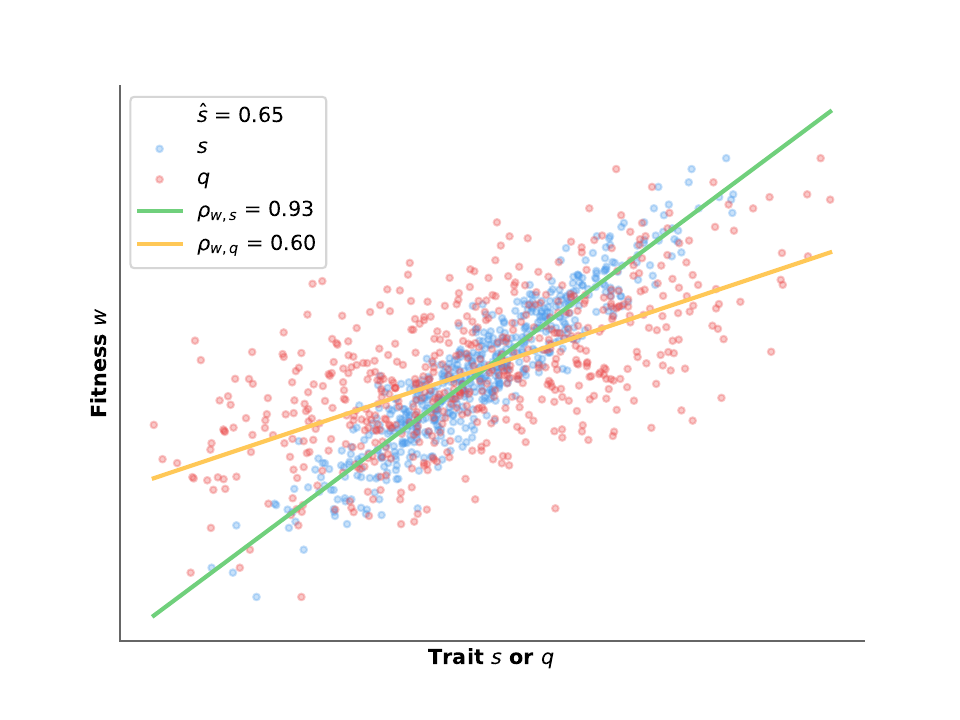}
            \caption{Fitness \(w\) versus $s$ and $q$}
            \label{fig:plot-wsq}
        \end{subfigure}

        \caption{
            Relations between traits within a single generation.\\
            \\
            (\subref{fig:plot-sq}) Green markers represent individual data points \((s, q)\), while the yellow line represents $\bar{\phi}(s)$. The higher the value of $\hat{s}$ is, the closer the points will be to the line.\\
            \\
            (\subref{fig:plot-wsq}) Scatter plot of fitness \(w\) versus $s$ and $q$ from simulation. The quality variable \(q\) is mapped onto the structural scale. Note that $\rho_{w,q} / \rho_{w,s} \approx \hat{s}$ under mediation.
        }

        \label{fig:plot-grid}
    \end{figure}

    \subsection{Multi-Generation Simulations}\label{subsec:muti-gen}

    On the page that follows, plots that demonstrate the relationship between mean structure, quality and fitness
    are plotted over a span of generations on the x-axis.

    \begin{itemize}
    \item In the simulation results shown, between 20 and 50 generations are depicted.
    \item The blue lines represent structural trait values.
    \item The error bars represent population variation in the trait.
    \item The red lines represent qualitative trait values.
    \item Smaller values of $\hat{s}$ will make the red and blue lines plot closer together.
    \item The  yellow line plots fitness.
    Higher values of fitness represent higher average fitness values within the overall population.
    \item All values are normalized to the same scale.
    \end{itemize}

    \begin{figure}[H]
        \centering

        \begin{subfigure}[b]{0.49\textwidth}
            \centering
            \includegraphics[width=\textwidth]{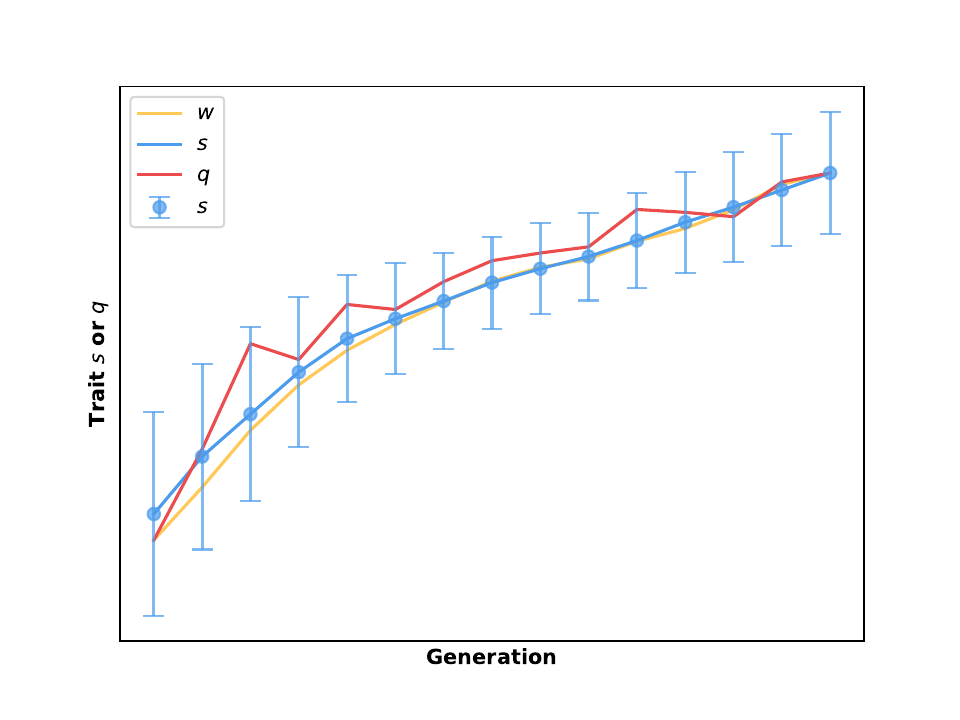}
            \caption{Natural Selection}
            \label{fig:plot-ns}
        \end{subfigure}
        \hfill
        \begin{subfigure}[b]{0.49\textwidth}
            \centering
            \includegraphics[width=\textwidth]{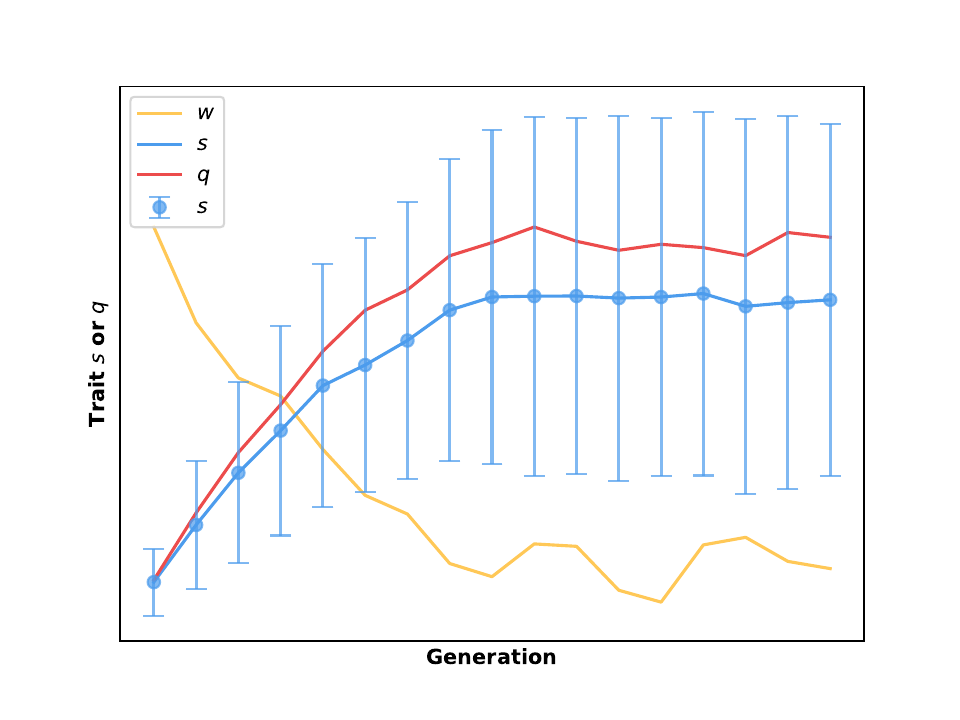}
            \caption{Overwhelmed By Mutation}
            \label{fig:plot-ns-inverse}
        \end{subfigure}

        \vspace{0.5cm}

        \begin{subfigure}[b]{0.49\textwidth}
            \centering
            \includegraphics[width=\textwidth]{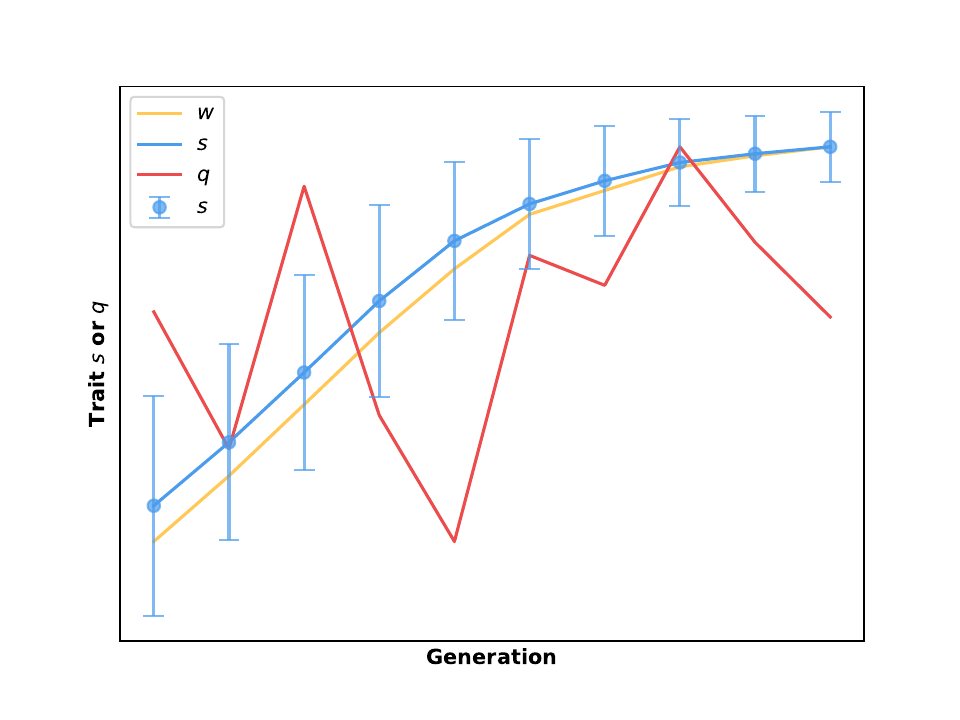}
            \caption{Exceeds QSL}
            \label{fig:plot-ns-qsl}
        \end{subfigure}
        \hfill
        \begin{subfigure}[b]{0.49\textwidth}
            \centering
            \includegraphics[width=\textwidth]{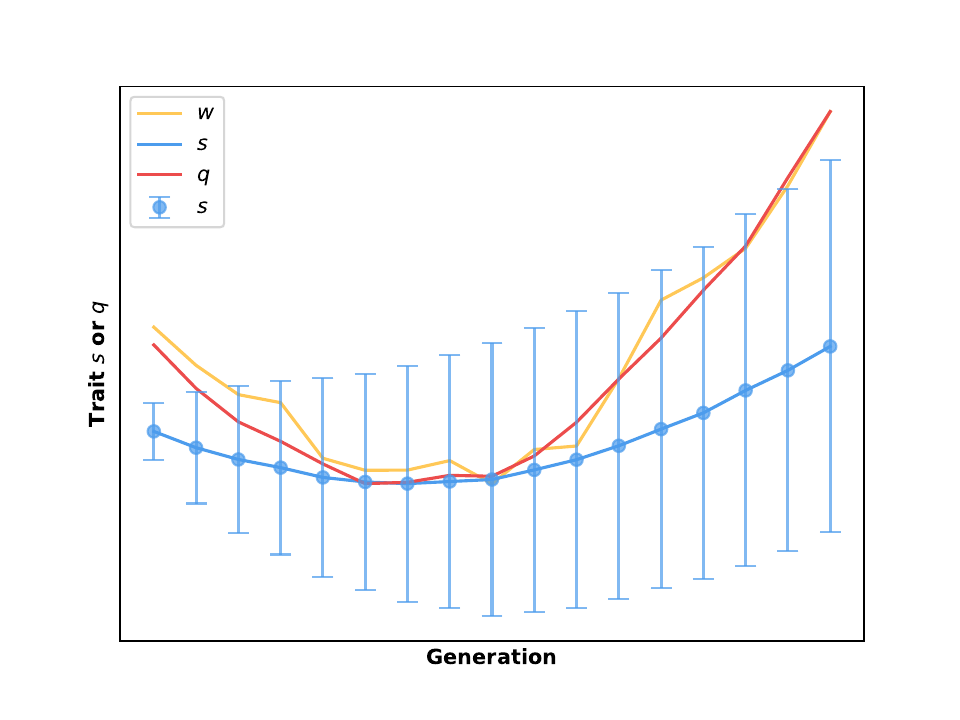}
            \caption{Rebound Selection}
            \label{fig:plot-ns-rebound}
        \end{subfigure}

        \caption{
            Generation by generation simulations.\\
            \\
            (\subref{fig:plot-ns}) This is effective natural selection under normal circumstances.
            You will notice all three lines plot closely together, and that there is a gradual reduction in variation due to the selective pressures taking hold.\\
            \\
            (\subref{fig:plot-ns-inverse}) This graphic shows an example of a system that is overwhelmed by transmission effects, such as mutation.
            Notice the increase in variation accompanied by a decrease in fitness.\\
            \\
            (\subref{fig:plot-ns-qsl}) This is a system under effective natural selection, but a small value of $\hat{s} = .006$
            leads to chaotic behavior in the qualitative realm.
            This is the behavior when \emph{qualitative selection limit} has been exceeded.
            Note that structural trait values increase gradually with fitness, while population variance decreases.\\
            \\
            (\subref{fig:plot-ns-rebound}) This is an interesting case with high transmission effects that are initially dominant, but are gradually overtaken by selective effects.
            Notice the initial slight upward trend in the trait value before reversing course.
        }
        \label{fig:ns-grid}
    \end{figure}

    \section{Appendix B: Higher-Order Examples} \label{sec:higher-examples}

    Two toy examples of higher-order systems are presented here for illustrative purposes.

    \subsection{A Conceptual Example}\label{subsec:conceptual-example}

    Imagine a person who knows no rules about the game of chess, except each piece is placed on a single square.
    If we show that person a piece, they will have no information to constrain their belief about which square the piece
    may reside on.
    So if we were to show them a pawn, their distribution would be uniform across the potential states $s_0$,
    each a number 1 through 64 representing a square on the board.
    \begin{equation*}\label{eq:chess-uniform}
    P_{\ell = 0}(s) = \frac1{64}
    \end{equation*}

    If we were to show the same pawn to a player with basic knowledge of the rules, they would know that a pawn can never
    appear on the first rank or the eighth rank.
    That player's perspective of the pawn's state on the board can be seen as a pair $(f, r)$ where $f$ is the file (given a letter $A$ through $H$)
    and $r$ is the rank (given a number 1 through 8).
    Each of the 64 squares on the chess board is named accordingly, e.g.\ A1, D4, etc.
    This information would allow them to constrain their distribution of beliefs about the location of the pawn:
    \begin{equation*}\label{eq:chess-informed}
    P_{\ell = 1}(f, r) = \begin{cases}
                             0 & \text{ when } r \in \{1, 8\},\\
                             \frac1{48} & \text{ otherwise}.
    \end{cases}
    \end{equation*}
    This higher-order knowledge gives them an informative prior over the states of $s_0$.

    A player with further knowledge of strategy and chess theory could be seen as having higher-order knowledge of $s_{2}$,
    which may be projected into a space that involves much more sophisticated states involving chunked patterns of pieces, etc.

    \subsection{A Simple Physical Example}\label{subsec:simple-physical-example}

    A simplistic version of a physical example can be seen in statistical mechanics.
    An ideal gas is fundamentally assumed to be a collection of particles where each particle holds some kinetic energy
    $s_0 = \{E_i\}$, drawn from a distribution, say
    \begin{equation*}\label{eq:particle-energy}
    E_i \sim \mathcal{N}(\bar{E}, \sigma),
    \end{equation*}
    where $\bar{E}$ represents the mean kinetic energy of a particle.

    Now we can take the perspective of the macrostate, where instead of considering individual particles, we can
    consider statistical properties of the set of particles.
    To take the most simplistic continuation of our example, we define our state to consist of only the temperature $s_1 = T$
    of the gas, which is determined in relation to its mean kinetic energy, $\bar{E}$ above:
    \begin{equation*}\label{eq:temperature}
    T = k_B \cdot \bar{E},
    \end{equation*}
    where $k_B$ is Boltzmann's constant to convert units from temperature to energy.

    Now we can see that while knowledge of $s_1$ does not give us any specific $E_i \in s_0$, it is a
    significantly informative prior over the distribution of $E_i$.

\end{document}